\documentclass[pra,twocolumn,amsfonts]{revtex4}
\usepackage{amsfonts}
\usepackage{amsmath}
\usepackage{bm}
\usepackage{graphics,graphicx}

\def\pr{{\textsl Phys. Rev. }}

\def\prl{{\textsl Phys. Rev. Lett. }}
\def\pr{{\textsl Phys. Rev.}}

\def\etal{{\textsl et al. }}

\newcommand{\beq}{\begin{equation}}
\newcommand{\eeq}{\end{equation}}
\newcommand{\beqa}{\begin{eqnarray}}
\newcommand{\eeqa}{\end{eqnarray}}

\newcommand{\ket}[1]{| #1    \rangle }

\begin{document}



\newcommand{\suba}[1]{_{_{A_{#1}}}}
\newcommand{\subat}[1]{_{_{\widetilde{A}_{#1}}}}
\newcommand{\subb}[1]{_{_{B_{#1}}}}
\newcommand{\subbt}[1]{_{_{\widetilde{B}_{#1}}}}
\newcommand{\subab}[1]{_{_{A_{#1}B_{#1}}}}
\newcommand{\subabt}[1]{_{_{\widetilde{A}_{#1}\widetilde{B}_{#1}}}}



\title{An AB effect without closing a loop }

\author{A. Retzker$^{1}$, Y. Aharonov$^{1,2}$, A. Botero$^{2,3}$, S. Nussinov$^{1}$, and B. Reznik$^{1}$ }

\affiliation{$^{1}$ Department of Physics
and Astronomy, Tel-Aviv University, Tel Aviv 69978, Israel}

\affiliation{$^{2}$ Department of Physics, University of Columbia, South Carolina}

\affiliation{$^{3}$ Departamento de F\'{\i}sica, Universidad de Los Andes,
    Apartado A\'ereo 4976, Bogot\'a, Colombia}

\date{\today}

\begin{abstract}

\bigskip
We discuss the consequences of the Aharonov-Bohm effect in setups
involving several charged particles, wherein none of the charged particles
encloses a closed loop around the magnetic flux. We show that in such setups,
the AB phase is encoded either in the {\em relative} phase of a bi-partite or multi-partite
entangled photons states, or alternatively, gives rise to an overall AB phase
that can be measured relative to another reference system.
These setups involve processes of annihilation or creation of electron/hole pairs.
We discuss the relevance of such effects in ``vacuum Birefringence" in QED,
and comment on their connection to other known effects.
\end{abstract}


\maketitle

\section{Introduction}
In the usual setup of the Aharonov-Bohm effect (AB) \cite{ab}, a
charged particle encircles a flux tube of total magnetic flux
$\Phi$, and collects the phase \beq \phi_{AB} = {e \over \hbar c}
\oint_{\it C}  \vec A \cdot\vec {dl} =\frac{e\Phi}{\bar h c}. \eeq
The AB phase, $\phi_{AB}$, has two important features. It depends
only on the topology of the trajectory via the winding number $n$.
Additionally the effect is "non-local"; at any intermediate point
along the trajectory, the magnetic fields vanish, and hence the
presence of the flux is locally undetectable. This is consistent
with the fact that the line integral of the vector-potential is
gauge invariant only along closed trajectories.

We shall discuss some new consequences of the AB effect in setups
involving several charged particles, none of which encloses a
complete loop around the flux. Under such circumstances, the AB
effect has new manifestations: the AB phase is encoded in the {\em
relative} phase of a bi-partite and multi-partite entangled state.
The AB topological non-locality gets transformed here into the
non-local property of the resulting entangled state.

Alternatively, the AB effect can give rise to an {\em overall}
phase of the system, which can be measured with respect to another
reference system. An electron-hole/positron pair is formed at one
location, and recombines at another location after encircling a
flux. The resulting photon then carries an overall AB phase.

In a related idea \cite{buttiker,buttiker1,buttiker2}, the AB phase
has been recently manifested in current-current correlations of
electrons in a Hanbury-Brown-Twiss interferometer. In this proposal
however, the effect is based on the indistinguishability of the
interfering electrons.

\begin{figure}
\begin{center}
\includegraphics[width=0.4\textwidth ]{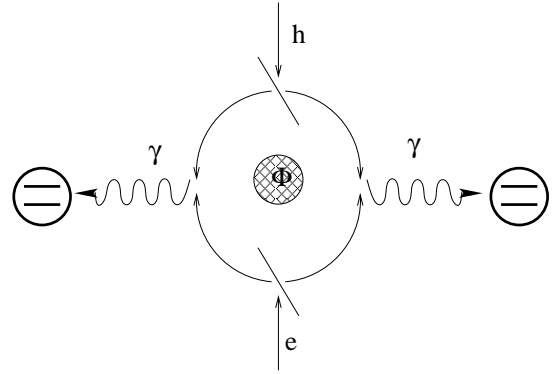}
\end{center}
\caption{The electron and the hole each completes half a loop and
annihilate to a photon. The photon can be absorbed by either the
left or the right atom. The bi-partite entangled atoms final state
depends on the AB phase. } \label{exp1}
\end{figure}

\section{Translating the AB phase to Entangled
states}\label{entangled}
Consider an electron and a hole that approach the fluxon from
opposite  upwards/downwards directions and pass through beam
splitters, as depicted in Fig. (\ref{exp1}) . The beam splitter
transforms the electron and the hole to a superposition of left and
right movers. The electron and hole can recombine into a photon
either on the left side or the right side. Adding up the phases
collected in each of the four parts of the circle in Fig.
\ref{exp1}, we find that the two  parts of the photon wave function
have a relative phase equal to the full AB phase. The postselected
state with no electron or hole, namely when a photon was created, is
then
\begin{equation}
\ket{1_L0_R} +e^{i\phi_{AB}}\ket{0_L1_R},
\end{equation}
where $\ket {n_L n_R}$ is the state with $n_L$ photons on the left
and $n_R$ photons on the right parts. Thus the flux becomes encoded
in the relative phase of a maximally entangled state.

It is instructive to compare between the usual measurement of the AB
phase in the standard setup and the present case. In our case, the
final photon state can be converted to a bi-partite entangled state
of a pair of two atoms.
${1\over\sqrt2}(|e,g\rangle+e^{i\phi_{AB}}|g,e\rangle$. The flux can
then be used to control "non-locally" the relative AB phase. This
phase can not be observed by performing measurements on only one
atom. It is manifested however in the correlations between the
results of the measurements performed locally on both atoms.

From the quantum information point of view, this setup provides an
interesting method to encode a classical bits into an entangled
state \cite{hiding}. For example, the observer that controls the
enclosed flux, can encode "0" in $\psi^+=
{1\over\sqrt2}(|e,g\rangle+|g,e\rangle$ and "1" in $\psi^+=
{1\over\sqrt2}(|e,g\rangle-|g,e\rangle$, by changing the enclosed
flux from $\Phi_0=0$ to $\Phi_1=h c/2e$.

\begin{figure}
\begin{center}
\includegraphics[width=0.25\textwidth ]{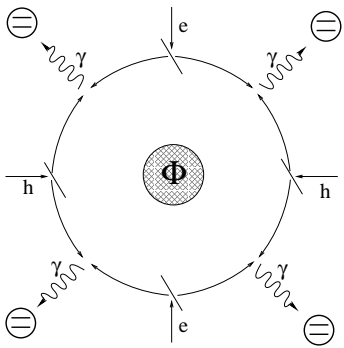}
\end{center}
\caption{The two electrons and the two holes each complete quarter
of a loop and together collect the topological phase which is
encoded in the phase of the emitted photons. The final state has
either two photons in arms 1 and 3, or in arms 2 and 4.}
\label{exp14}
\end{figure}

The above scheme can be extended to $n$ electrons and $n$ holes.
For example, in Fig. \ref{exp14}, two electrons and two holes
approach the flux from four different directions. If a pair
recombines then the two neighboring pairs cannot recombine, thus
either two opposite photons are emitted or the other two opposite
photons are emitted. The output state is then $\vert 1010
\rangle+e^{i\phi{AB}} \vert 0101\rangle$, where $0$ and $1$
designate the fock state of the four output channels of the
photons.


In order to compute the resulting states in the above and similar
setups, we make the following assumptions.  The dynamical evolution
of the system at the creation and annihilation vertexes can be
obtained by applying creation-annihilations operators to the wave
function. In particular, in the vertex where a photon creates an
electron hole pair, or when an electron hole creates a photon, we
have $a_{photon}\, a_e^\dagger\,\, a_h^\dagger$, and $a_e\,\, a_h
a_{photon}^\dagger$, respectively. In these process, the net energy
and momentum exchange of the charged particle with with the matter
can be made small enough, so that the coherence of the process is
maintained. In order to calculate the effect of the AB flux on the
different charged particles trajectories, it is useful to use a
particular gauge. In the singular gauge, the vector potential
vanishes except along a singular line that emanates from the fluxon.
In this gauge, only a charged particle that crosses the line
accumulates a phase. For example, $a_e^\dagger\rightarrow
a_e^\dagger e^{-i \phi }$, $a_h^\dagger\rightarrow a_h^\dagger e^{i
\phi }$.

\section{AB effect with photons}\label{photons}
\begin{figure}
\begin{center}
\includegraphics[width=0.4\textwidth ]{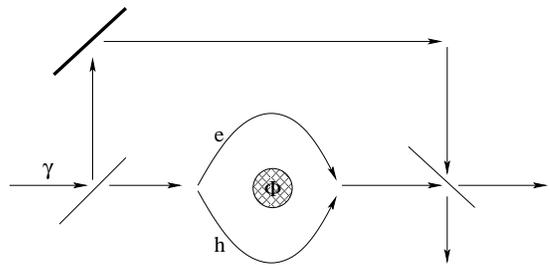}
\end{center}
\caption{The photon creates an electron and a hole and collects
the topological phase on recombination. The process takes place in
one of the arms of a Mach-Zender interferometer.} \label{exp2}
\end{figure}

In a different variant  the AB phase is transferred to  a  photon
as depicted in Fig. \ref{exp2}. The photon creates an electron and
a hole. The latter can move on both sides of the flux and then
annihilate back into a photon carrying the topological phase. The
symmetry between electron-up, hole-down and electron-down, hole-up
could be broken using an external electric field. To measure this
phase we need  the two beam splitters and mirror shown in Fig.
\ref{exp2} yielding two alternative paths for the photon, only one
of which is affected by the AB phase. This yields a final output
signal with small modulations periodic in the flux due to
interference.

\section{Vacuum birefringence}\label{birefringence}
The well known  "Vacuum Birefringence" in QED can be related to the
above set-ups. Thus (PVLAS experiment \cite{pvlas}) envision laser
light propagating in vacuum along the z direction say in a
transverse B field. The virtual electron box diagram generates for
the low energy ($E_{\gamma} \ll m_e c^2 $) an effective Euler
Heisenberg Lagrangian \cite{euler,schwinger,adler,bialynicka}:
(in naturalized Gaussian units $\hbar=c=1$)
\begin{equation}
L_{eff}= \frac{2\alpha^2}{45(4\pi)^2m^4} \left[ (E^2- B^2)^2 + 7(E
\cdot B)^2 \right]
\end{equation}
with the four E,B factors representing external fields and or
photons. $L_{eff}$ generates in particular "Vacuum Biefringing",
namely a relative phase between the x and y polarizations of the
photon. Can we heuristically understand this in a manner emphasizing
the role of an AB type phase? The photon can virtually convert into
an electron-positron pair which after free propagation (in
configuration space) recombine back into the original photon. If a B
flux threads the path closed jointly by the electron and positron
then the amplitude picks up an extra AB phase. For the case of a
uniform B field $\Phi_{AB} = AB sin \theta$ with $\theta$ the angle
between the "plane" where $e^+ e^-$ move and B with A the "net"
signed area enclosed. During their life-time $\Delta t
=\frac{h}{2mc^2}$, the electron/positron travel distances $l =
\frac{h}{mc}$ and $A = l^2 = \left(\frac{h}{mc}\right)^2$. The
"dipole" interaction $ E _{photon}.(p^+-p^-)$ tends to create/
annihilate the pair in the polarization plane of the photon. The
amplitude of polarization perpendicular to the B field picks an AB
phase relative to the other, orthogonal, polarization adding a small
circular polarization to the primarily linearly polarized light or a
small "ellipticity". The above argument fails however. The explicit
Euler Heisenberg effective Lagrangian and more generally Furry's
theorem \cite{itzykson,weinberg} forbid trilinear photon coupling or
polarization changes linear in B external. Indeed for every loop
traversed by the $e^+$ and $e^-$ in a given sense there is loop of
equal amplitude with $e^+ \leftrightarrow e^- $ traversed and the
opposite AB phase! A finite effect ensues in next order: The $U =
\mu\cdot B$ interaction with the loops'  magnetic moment: $\mu =
eA\omega = eAmc^2/h$ enhances the probability amplitude of one
orientation of the loop relative to the other by $~ U/mc^2$ i.e by
$BA \approx B (h/mc)^2$ avoiding the above cancellation of the AB
phases and yielding a net effect of the correct form $~ e^4 B^2
/m^4$.

\section{Some comments on the "AB Zeeman" and "AB
Faraday Effect"}\label{comments}

In sections \ref{entangled},\ref{photons} above new variants of the
AB effect where no single charge particle encircles the flux were
utilized to transfer the AB phase to photons. Here we note that
aloso the conventional AB effect can yield such a phase transfer via
an "AB Faraday" effect.
For extended B fields the "classical", non-anomalous Zeeman and
Faraday effects are well understood: The $\vec{\mu} \cdot \vec{B}
\propto g\, \vec{l}\cdot\vec{B}$ interaction splits the $m$
sublevels. The resonant absorption frequency of left and right
circularly polarized light separate by 2gB and the corresponding
indices of refraction differ accordingly by:$ n_R-n_L \propto
gB_{||}/\omega$, with $B_{||}$ - the (say z) component of $B$
parallel to the light propagation. Along a distance $L$ this does in
turn rotate the initial plane (of the linearly polarized light) by:
$(n_R-n_L)\cdot L/\lambda$ ($\lambda$ is the wave-length divided by
$2\pi$.)

In the AB effect the electron picks up the magnetic field induced
phase despite being at all times in the B=0 region. The "AB Zeeman"
effect is the energy shift of such a particle. Let a cylindrical
shell of inner / outer radii r,R be threaded by a flux $\phi$ along
its(z) axis. The states of non-interacting electrons in this shell
where a cylindrically symmetric potential exists $\Psi_{k,m,n}$ have
energies $E_{k,m,n}$ depending on the z components of the linear and
angular momenta $k=p_z/h,m=l_z/h$ and a remaining "radial" quantum
number $n$. The introduction of the flux shifts changes the angular
momentum quantum number:$l_z'=l_z-\alpha$. It leaves the single
valued wave functions changing the energies via the substitution
$E_{k,n,m} \rightarrow E_{k,n,m+\alpha}$. The AB flux is a "modular"
variable and the levels cross for $|\alpha|=1/2$, hence the shift
above is by the smaller of non integer part of the flux or its
complement to an integer.

The Zeeman shift is linear in B (for "small" B ) and is not
periodic. The AB Zeeman effect {\it is} periodic in the flux and,
for fixed area of fluxons, in the field B. When the field is uniform
and the sample continuous we find that the AB shift becomes the
Zeeman effect. It is interesting to note that in a cylindrical
sample with mobile electrons (and/or holes) with a common (small)
radius a, the levels corresponding to paths enclosing the hole with
radii peaked near r=a, and a periodic dependence on B, See \footnote
{The above expression for the flux induced energy shift depends on
the azimuthal ($\theta$) symmetry when $l_z$ is a good quantum
number. A strong $\theta$ dependent potential breaking this symmetry
can generate, bound states localized in $\theta$. Only the small
overlap of the left and right tails of the wave function allows such
states to enclose the fluxon and the energy shift is accordingly
reduced.}.

This is illustrated in excitonic states at the rim of the holes bound by modified coulomb
$1/|\theta-\theta '|$ potential. This is equivalent to looking for bound
states in the one dimensional problem on an interval $[0,2\pi a]$ where for
say the even parity sector we demand that $\psi '(0)=\psi '(2\pi a)=0$ which
can be solved with and without the fluxon the introduction of which
changes the $\frac{d}{id\theta}$  into $\frac{d}{id\theta}-\frac{\alpha}{2\pi}$\cite{romer}.

The second setup (fig. \ref{exp2}) can be manifested in a photon
exciton system \cite{yamamoto} wherein angular momentum conservation
simplify the calculations. In semiconductors when a photon creates
an exciton a $R$( photon creates a $X_+$ exciton and a $L$ photon
creates a $X_-$ exciton(where $L$,$R$ are orthogonal circular
polarizations and $X_-$ and $X_+$ are orthogonal state of the
exciton with different angular momentum). Each exciton collects the
phase with a different sign since in the relative coordinates the
charge rotate in a different direction. Thus for the proper choice
of flux the relative phase between $X_+$ and $X_-$ is $-$. Since the
angular momentum is in the direction of the propagation of the
photon, in this setup the magnetic field should be parallel to the
momentum of the photon. Hence a photon with polarization in the $x$
direction would change polarization to the $y$ direction. In this
scheme the AB phase is manifested in the rotation of polarization.

The AB Faraday effect is the rotation of polarization plane for
light propagating in the z direction i.e., along the fluxon and
axis of the cylindrical sample. For the exciton the idea is the
same as for the normal effect except that the energy shifts are of
the exciton and not of the electron. The energy shift explanation
is valid for weak magnetic fields, for stronger magnetic fields
 the reason for the rotation is analogous to the explanation of
the AB rotation of polarization. To avoid a strong decline of the
effect with the distance from the above axis/fluxon, the wave length
of the light can be of order the radius of the cylinder.

In order to estimate the magnitude of the effect we calculate the
regular AB Faraday effect for a charged particle condensate
constrained to a narrow ring, with a flux passing through the ring
axis.  Let $\phi$ be the angle on the ring, $\psi(\phi,t)$ be the
condensate wave function on the ring, and assume the form
\begin{equation}
\psi(\phi,t) = \sqrt{n}e^{\frac{i}{h}S(\phi,t)}\,
\end{equation}
where $n$  is a constant particle density, integrating to    a total
number of particles $N$ on the ring. We assume a flux $\Phi = \beta
\Phi_o$  along the $z$  axis, and an  incident circularly-polarized
electromagnetic plane wave along the same direction
\begin{equation}
\vec{A}_{inc} = A_\pm \hat{\varepsilon}_\pm e^{i (k z - \omega t)
}\, .
\end{equation}
where $\hat{\varepsilon}_\pm = \frac{1}{\sqrt{2}}\left[\hat{x} \pm
i \hat{y} \right]$. In a low density approximation, the phase
satisfies the Hamilton-Jacobi equation
\begin{equation}
-\dot{S} = \frac{1}{2 m }\left( \frac{1}{R}\frac{\partial S
}{\partial \phi} - \frac{e}{c}A_\phi \right)^2 \,
\end{equation}
where $A_\phi$ is the $\phi$-component on the ring of the total
vector potential  $\vec{A}_{tot} = \frac{1}{2 \pi R} \beta \Phi_o
\hat{\phi} + \vec{A}_{inc} \, $. Note that $\frac{e}{2 \pi
c}\Phi_o = \beta \hbar$, so that, assuming linear response to
 $A_{inc}$, we get  response  components $S(\phi,t) = S_{\pm}e^{-i \omega t \pm i \phi
 t}$,
\begin{equation}
S_\pm(\omega) = \frac{1}{\sqrt{2}}\left(\frac{e R}{c}\right)
\frac{ \beta \omega_o}{\beta \omega_o \pm \omega}A_\pm \, ,
\end{equation}
with $w_0=\frac{\hbar}{2mR^2}$. The associated current densities on
the ring are
\begin{equation}
\vec{J}_{\pm}  =  \pm i \frac{n e^2}{m
c}\left(\frac{\omega}{\omega \mp \beta \omega_o}\right)
\frac{A_\pm}{\sqrt{2}} \hat{\phi}\,  .
\end{equation}
The scattered fields preserve the incident polarization, resulting
in a forward scattering amplitudes $f_\pm(\Theta=0)=\
\frac{N}{2}\left(\frac{ e^2}{m c^2}\right)\left(\frac{\omega}{\omega
\mp \beta \omega_o}\right)$. Designating by $r_0=\frac{e^2}{mc^2}$
the "classical" radius of the electron, using the dimensionless s
matrix
\begin{equation} S_\pm=1+if_\pm k=1+\frac{i r_0
\omega}{\lambda(\omega \pm \beta \omega_0)}
\end{equation}
and specializing to the limit case of $N=1$ i.e. a single electron
in the ring, we finally find a rotation angle of order $\Delta
\theta =\Delta S=\frac{r_0
\beta}{R^2}\frac{\hbar}{mc}\approx\frac{\beta 10^{-24}cm^3}{R^2}$
for $R=\lambda\approx 10^{-4}cm$ the angle is very tiny $\Delta
\theta \approx 10^{-15}$

In conclusion, we have discussed some new features of the AB
effect, and showed that the AB phase can manifest itself without any loops
being closed by a single particle. We have discussed several variations of this idea,
and showed that the nonlocal AB phase can be stored either in an entangled
bi-partite or multi-partite states, or in the overall phase of photons
or in the direction of polarization.

\begin{acknowledgments}

We would like to thank A. Casher, M. Heiblum, I. Neder  and L.
Vaidman for helpful discussions.

\end{acknowledgments}

\end{document}